\documentclass[12pt,graphics]{iopart}

\newcommand{\Journal}[4]{{#1} #4 {\bf #2}, #3 }
\usepackage{iopams}  
\usepackage{graphicx}
\usepackage{color}
\usepackage{float}
\usepackage{textcomp}
\usepackage{amssymb}
\usepackage{graphicx}
\usepackage{setspace}
\usepackage{esint}
\usepackage{enumitem}
\usepackage{url}
\graphicspath{{.//}{figures//}}

\newcommand{\NIMA}{{\em Nucl. Instrum. Methods} A}
\newcommand{\NIMB}{{\em Nucl. Instrum. Methods} B}

\newcommand{\NPA}{{\em Nucl. Phys.} A}

\newcommand{\PRC}{{\em Phys. Rev.} C}
\newcommand{\CPC}{{\em Chin. Phys.} C}

\newcommand{\PRP}{{\em Phys. Rep. }}

\newcommand{\NAT}{\em Nature}

\newcommand{\bnel}{\mbox{$\bar{\nu}_e$} }

\newcommand{\obb}{0\mbox{$\nu\beta\beta$ - decay} } 
\newcommand{\zbb}{2\mbox{$\nu\beta\beta$ - decay} }

\newcommand{\gess}{\mbox{$^{76}$Ge }}

\newcommand{\tinhz}{\mbox{$^{112}$Sn }}

\newcommand{\tinhv}{\mbox{$^{124}$Sn }}
\newcommand{\tehv}{\mbox{$^{124}$Te }}

\newcommand{\xehs}{\mbox{$^{136}$Xe }}

\newcommand{\tzn}{\mbox{T$_{1/2}^{2\nu}$} }

\newcommand{\be}{\begin{equation}}
\newcommand{\ee}{\end{equation}}
\def\bea{\begin{eqnarray}} 
\def\eea{\end{eqnarray}} 
\newcommand{\ra}{\rightarrow }

\begin{document}

\today

\title{A tin-loaded liquid scintillator approach for the 2 neutrino double-beta decay measurement of $^{124}$Sn}
%Possibility of measuring long-living \zbb emitters in the presence of short-living ones or }

\author{O. Chkvorets$^a$, C. Kraus$^a$, J. K\"uttler$^b$, V. Lozza$^c$, B. von Krosigk$^d$ and K. Zuber$^b$}

\address{$^a$Laurentian University, 935 Ramsey Lake Road, Sudbury, ON P3E 2C6, Canada}

\address{$^b$Technische Universit\"at Dresden,Institut f\"ur Kern- und Teilchenphysik,
Zellescher Weg 19, 01069 Dresden, Germany}

\address{$^c$Universidade de Lisboa, Departamento de Fisica, Faculdade de Ciencias, Campo Grande, Edificio C8, 1749-016 Lisboa,
Portugal}

\address{$^d$The University of British Columbia, Department of Physics and Astronomy, Vancouver, BC, Canada}

\ead{zuber@physik.tu-dresden.de}

\begin{abstract}
A new experiment based on tin-loaded scintillator is proposed to measure the \zbb half-life of \tinhv for the first time. 
Measurements of long term stabilitiy and optical properties of the produced scintillator are presented. In addition a sophisticated estimation of the background
due to cosmic ray spallation on tin has been performed. It is shown that such a measurement is feasible and can reach half-lives
of 10$^{22}$ years, thus covering all current theoretical predictions for this decay mode.
%Also \tinun does double-beta decay, but has a very small Q-value of about 300 keV. 
%Furthermore,  the $\beta^+$/EC-decay potential is also discussed. An outlook on a potential neutrino-less
%decay search for \tinhv is given. 

%of However \tinhz could be interesting as beta+/EC with a Q-value of 1922 keV.

\end{abstract}

\maketitle

\section{Introduction}

The search for physics beyond the standard model (BSM) is a wide spread
activity in accelerator and non-accelerator particle physics, because the Standard Model
as successful as it is cannot be the final theory. As an example, the Standard Model does
not provide a particle candidate for dark matter. Furthermore, conservation rules like baryon
and lepton number are phenomenological and have no known theoretical background like charge
conservation. Hence, various processes are explored like proton decay, neutron-antineutron oscillations, 
charged lepton flavour violation and more. In this context also the total lepton number $L$ plays 
an important role. The golden channel to search for an actual total lepton number violation is the
neutrino-less double-beta decay of a nucleus (A,Z)
\be
(Z,A) \ra (Z+2,A) + 2 e^-  \quad (\obb).
\ee   
This nuclear decay mode violates lepton number by 2 units and would also prove that neutrinos are Majorana particles.
To observe this process, single beta decay has to be forbidden by energy conservation or at least strongly suppressed due
to a large change of the involved spins. For this reason only 35 potential double-beta emitters exist. As the phase space for these decays
scales strongly with the Q-value (in case of \obb with $Q^5$ and in case of \zbb with $Q^{11}$), experimental searches are typically using only
 nuclides with a Q-value above
2 MeV, which is reducing the list of candidates to 11. 
Lower limits on half-lives well beyond 10$^{25}$ years of the neutrino-less mode have been measured for the isotopes \gess and \xehs  \cite{ago17,alb14, KL}. \\
In addition, the allowed  process of nuclear double-beta decay 
\be
(Z,A) \ra (Z+2,A) + 2 e^-  + 2 \bnel \quad (\zbb)
\ee      
will occur. It is the rarest decay measured in nature and has been observed 
in about ten isotopes.
The following will focus on this mode. The \zbb measurements are interesting by itself in several ways: 
First of all, this process is free of any unknown particle physics quantities like effective
Majorana neutrino masses and other Beyond Standard Model physics appearing in neutrino-less double-beta decay. 
Hence the decay rate is just linked to a phase space $G^{2\nu}$ and a nuclear matrix element $M^{2\nu}_{GT} $
\be
\left(\tzn\right)^{-1} = G^{2\nu} \times \mid M^{2\nu}_{GT} \mid^2\,.
\ee
All double beta decay transitions to the ground state of the daughter 
are of type $0^+\rightarrow 0^+$, where the intermediate states will be $1^+$. Hence, transitions from the double-beta emitter to
the intermediate $1^+$ states are of Gamow-Teller type and  their strength
has been measured using ($^3$He,t) reactions for some of the double-beta candidates. 
From the nuclear structure point  of view \zbb is interesting because various approaches to determine the matrix elements can be
compared with measured half-lives. 
Also very appealing is a comparison of double-beta
emitters from the same element 
(separated by a difference of two neutrons in the same shell).
Nature offers a few elements  with 2(3) emitters:  $^{46,48}$Ca, $^{80,82}$Se,
$^{94,96}$Zr, $^{98,100}$Mo, $^{114,116}$Cd, $^{122,124}$Sn, $^{134,136}$Xe and $^{146,148,150}$Nd.
However, experiments using one of these elements are always dominated by the 
nuclide with the highest Q-value, which will make a measurement of the remaining other nuclides
almost impossible unless isotopical modifications (enrichment) are performed.\\
Relatively unexplored double-beta nuclides in this respect are the tin isotopes. The Q-values for $^{124}$Sn and $^{122}$Sn are 2292.64 $\pm$ 0.39 keV and $373.1 \pm 2.7$ keV respectively \cite{ame17}, indicating that the \tinhv decay will be by far the dominant mode. Experimental lower half-life limits from searches into excited
states of \tehv exist \cite{daw08,daw08a,bar08}. A lower half-life limit for the neutrino-less
decay has been given in \cite{hwa07,hwa09}. Activities to build a cryo-bolometer system using tin as
absorber have been started \cite{vivek}. \\
On the proton rich side of tin isotopes there is \tinhz, which allows for double electron capture (EC) and
$\beta^+$/EC-decay with a Q-value of 1922 keV. Again lower limits for excited state transitions exist \cite{daw08,daw08a,bar08}. \\
Various calculations of \zbb half-lives for \tinhv exist. Half-life values given  (in units of $\times 10^{20}$years)
are 0.78 \cite{suh98}, 1.3/0.43 \cite{suh11}, 1.8 \cite{sim13} and 2.9 \cite{cau99}, respectively.
However, a recent shell model calculation \cite{hor16} claims almost an order of magnitude longer half-life of  1.6 $\times 10^{21}$ years \cite{hor16}. Furthermore, new studies taking into account the quenching of the axial-vector constant $g_A$ are predicting
half-lives in this region as well \cite{suh15}.\\
Here we focus on a potential measurement of the \zbb of \tinhv based on tin-loaded scintillators.
% but also will explore the sensitivity for \tinhz. 
The natural abundance of \tinhv is 5.79\%, isotopical enrichment to more than 90\% is feasible. With the given new theoretical 
estimates of the half-life
we design an experiment which will have a sensitivity for \zbb beyond all current calculations.\\

%\section{Experimental setup\textcolor{red}{All}

%\pagebreak 
\section{Tin-loaded scintillator and purification}

Tin-loaded liquid scintillators are known as high-Z liquid scintillators \cite{Knoll} and for double-beta decay searches \cite{hwa07,hwa09}.
In a very first test tin was loaded in linear alkyl benzene (LAB) - a perfect scintillator for large scale detectors planned for example in 
the SNO+ and JUNO experiments.
The absorption and emission spectra of tin-loaded LAB are shown in Fig. 1 and 2.
% %
\begin{figure}[hhh]
\label{fig:abs}
\centering
\includegraphics[width=0.75\columnwidth]{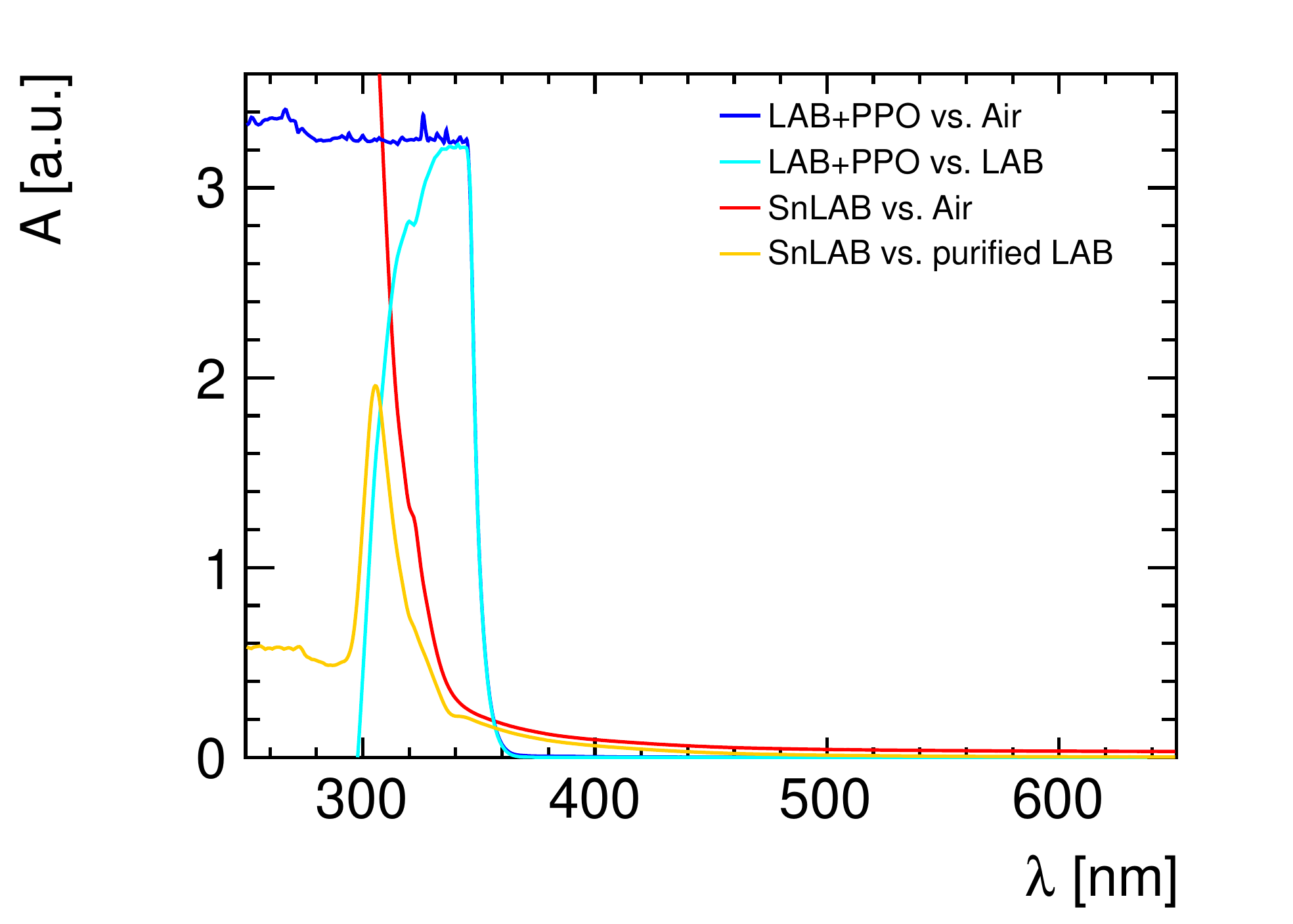}
\caption{Absorption spectra of pure and Sn-loaded  (SnLAB) scintillator as a function
of wavelength. For one measurement air (dark blue) and for a second specially purified LAB (blue) was used as reference.
 There are no tin-induced absorption features in the region from about 360-600 nm, which shows the
 feasibility of the proposed experiment in terms of optics. The attempt to 
 measure the Sn-contribution individually (yellow) created a fake feature at 310 nm in the measurement.
 The reason is that here purified LAB was measured relative to non-purified LAB, which creates
 different absorption spectra. The rise in absorption around 360 nm (blue)  is due to the PPO component in the scintillator.}
%\label{pic:sensi}
\end{figure}
% % %
\begin{figure}[h]
\label{fig:emi}
\centering
\includegraphics[width=0.75\columnwidth]{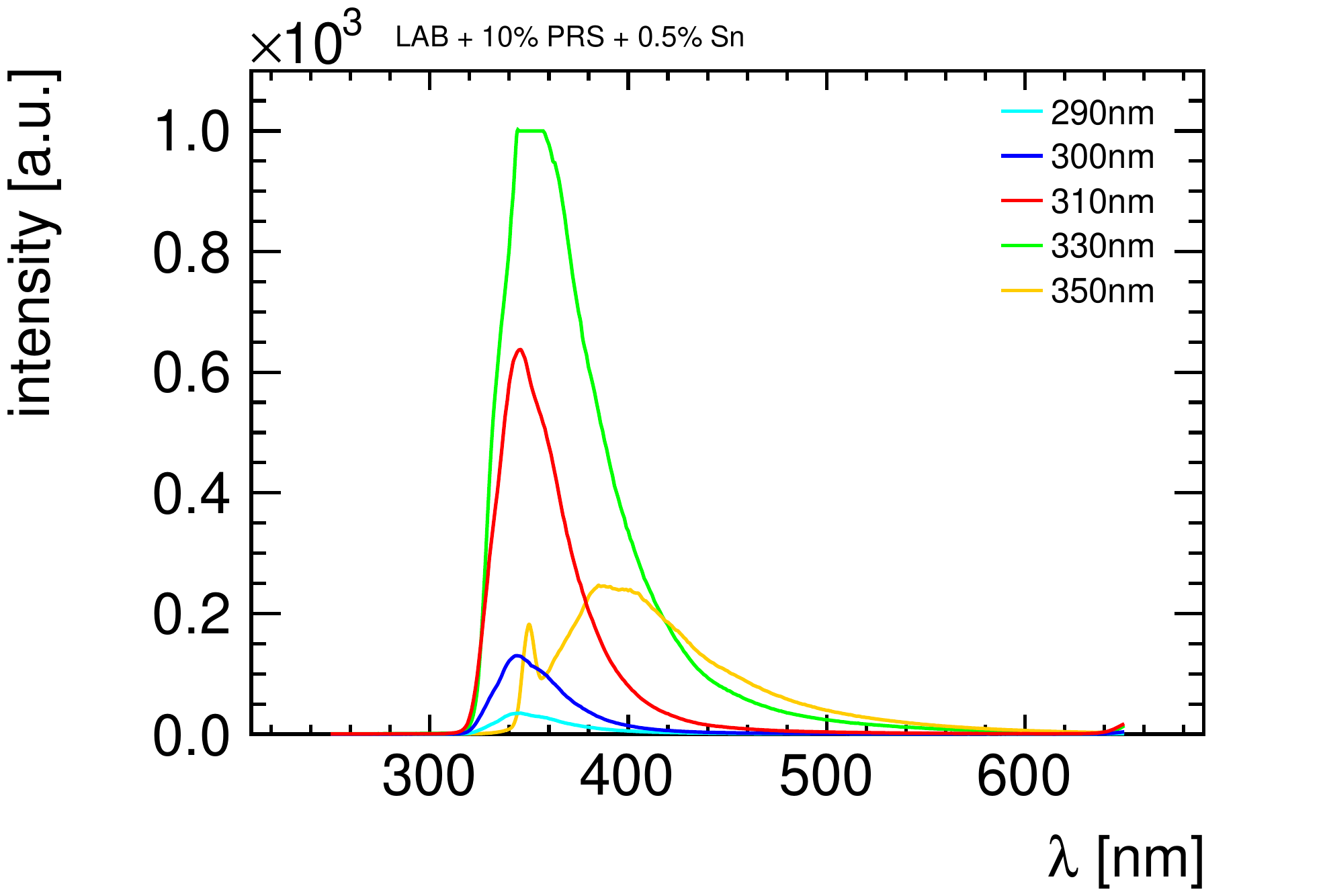}
\caption{Intensity of the emission spectrum of 0.5\% loaded SnLAB as a function of wavelength. The plateau at the 330 nm spectrum (green) is
a saturation effect of the apparatus. PRS is a surfactant used to load the scintillator.}
%\label{pic:sensi}
\end{figure}

There are several organometallic chemicals containing tin which were tested with various liquid scintillators. One of them - tetrabutyltin (TBSn, Fig. 3) - is found to be soluble in LAB. This is the first test of TBSn loaded in LAB as for the authors knowledge. The major physical parameters for the tin-loaded liquid scintillator are light yield, optical transparency and stability over a period of several years. Another important technical parameter is the flash point temperature. TBSn has a high enough flash point for safety requirements, as the experiment has to be performed underground, which requires special care. 
The physical and chemical properties of the TBSn are listed in Table 1.\\
%Table \ref{TBSn-properties} 

\begin{figure}
\label{fig:TBSn}
\begin{center}
\includegraphics[width=0.5\linewidth]{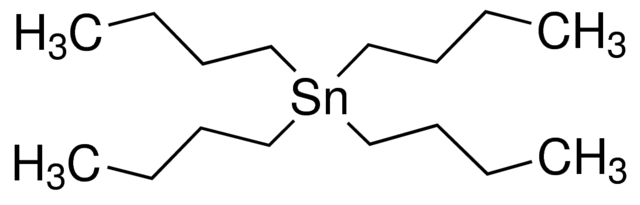} 
\includegraphics[width=0.4\linewidth]{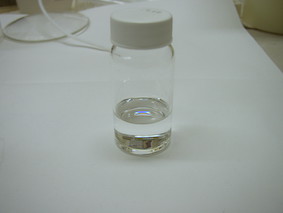}
\caption{Left: Chemical structure of tetrabutyltin (TBSn). Right: TBSn sample. }
\end{center}
\end{figure}
%\begin{figure}

%\caption{Tetrabutyltin (TBSn)}
%\label{fig:TBSn}
%\end{figure}

\begin{table}
\label{Tab1}
\begin{center}
\begin{tabular}{|c|c|}
\hline
\hline \rule[-2ex]{0pt}{5.5ex} Chemical Formula & $\rm{ C_{16}H_{36}Sn}$\\ 
\hline \rule[-2ex]{0pt}{5.5ex} Molecular Weight &  347.17\\ 
\hline \rule[-2ex]{0pt}{5.5ex} Density g/ml & 1.057 \\ 
\hline \rule[-2ex]{0pt}{5.5ex} Boiling point,10 mmHg, C & 127-145 \\ 
\hline \rule[-2ex]{0pt}{5.5ex} Flash point, C & 107\\ 
\hline \rule[-2ex]{0pt}{5.5ex} tin content w/w, \% & 34.2 \\ 
\hline 
\end{tabular}
\end{center}
\caption{Various properties of tetrabutyltin (TBSn).}
\end{table}

\subsection{Relative light yield of tin-loaded LAB vs. concentration}

Sigma Aldrich tetrabutyltin (93\%, technical grade) was used for tests presented in this paper. Distilled LAB was mixed with 4g/l PPO used as fluor. TBSn was mixed with LAB-PPO in five 25 ml scintillation vials with 0.5\% 1.5\%, 3.0\%, 5.0\% and 10\% mass content of tin, respectively.
 One vial with pure LAB-PPO was prepared for reference and stored together with the tin-loaded vials. In order to improve the optical properties of the cocktails, dry nitrogen gas was purged in each vial for 20 minutes to remove the oxygen molecules. The vials were closed and stored in a dark place 
at room temperature during a period from November 2015 to May 2017. The relative light yield measurements were performed in the beginning of this period and in the end to verify stability of the TBSn-scintillator. The same vial with pure LAB-PPO was used as reference.
The samples were placed on top of a  PMT using vacuum grease for optical coupling. 
A $^{137}$Cs source was then placed in close contact with the vial 
%(how do you verify the reproducibility of the position? was there a special holder for the source?) 
and the spectra were collected using a standard ORTEC MCA. The relative yield was calculated as the ratio of the Compton edge amplitudes extracted from the tin-loaded scintillator spectrum and the LAB-PPO one \cite{tinLS-TAUP2017}. The measured spectra with the $^{137}$Cs source of the LAB-PPO and Sn-loaded LAB are shown in Fig. 4.
In Table 2 the relative light yields for different concentrations are presented.\\

\begin{table}
\label{Tab2}
\begin{center}
\begin{tabular}{|clc|c|}
\hline  & November 2015 & May 2017 \\ 
\hline tin Concentration,\%	& Relative LY,\%& Relative LY,\%\\
\hline  LAB-PPO	&	100& 100\\
\hline 0.5 &  96 & 95 \\ 
\hline 1.5 & 85 & 82 \\ 
\hline 3.0 & 77 & 75 \\ 
\hline 5.0 & 67 & 68 \\ 
\hline 10.0 & 56 & 55 \\ 
\hline
\end{tabular}
\end{center}
\caption{Light yield measurements using the Compton edge of a $^{137}$Cs source with a time gap between the measurements of about 1.5 years. Shown are the values for various tin concentrations. No statistical significant change has been seen.}
\end{table} 

\begin{figure}
\centering
\includegraphics[width=0.7\linewidth]{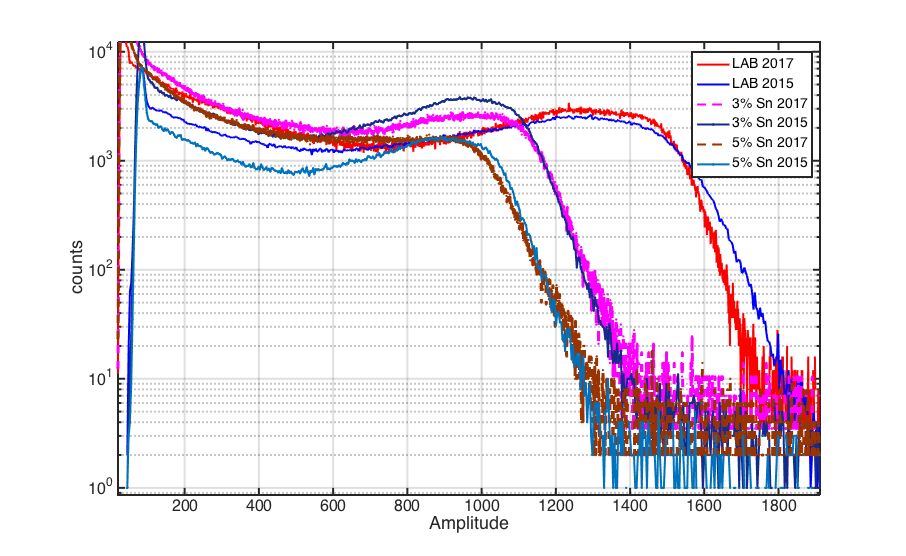}
\caption{$^{137}$Cs spectra are taken with LAB-PPO and TBSn+LAB (3\% and 5\% of tin) in November 2015 and May 2017. The TBSn liquid scintillator is stable over this period.}
\label{fig:Spectra-LAB-TBSn-2015-17}
\end{figure}

%The measured spectra are presented in Fig.\ref{TBSn_spectra}

\subsection{TBSn Purification}
The primary method of TBSn purification is via distillation. Distillation has a proven high removal efficiency of intrinsic U and Th contamination in TBSn. Additionally, it removes the majority of the nuclides produced by cosmogenic neutron and proton activation of the Sn itself (see Section 3) which are not chemically bonded to the butyl chains. Fast neutron and proton activation on Earth surface will likely break the butyl-tin bonds such that the produced nuclei are not part of the TBSn molecule anymore and can be removed by distillation. On the other hand, this is most likely not the case for nuclides produced by thermal neutron capture of tin isotopes. In this case the tin compound remains in the same chemical form. Differently from the fast neutron activation, thermal neutron capture can also occur deep underground (UG). Depending on the neutron flux measured at the experimental location a reduction of the thermal neutron flux might be necessary, such as placing the detector far from the rock wall or using neutron absorbers.\\
The purification via distillation has also the advantage to significantly improve the optical properties of the TBSn cocktail, as shown in 
Fig.5. 
Another purification method is based on liquid-liquid extraction. TBSn is, in fact, soluble in LAB while both of them are not soluble in water. The liquid-liquid extraction method is effective in removing ionic impurities, like the radon daughters -- lead, bismuth, and polonium --, and cosmogenic metal impurities. A big advantage of the liquid-liquid-extraction method is the possibility of using it for online purification of the cocktail during the data taking period.
\begin{figure}
\centering
\includegraphics[width=0.7\linewidth]{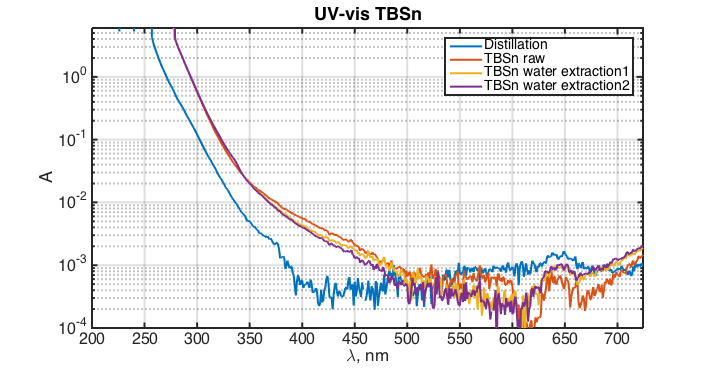}
\caption{Effect of purification methods on TBSn absorbance. Distillation significantly improves the optical properties of TBSn. Measurements were performed with a 1 cm cell. }
\label{fig:UV-vis-TBSn}
\end{figure}
%\newpage

 \section{Cosmogenic background}

A potential background for all double-beta decay experiments are the nuclides produced by the cosmic neutron and proton activation of the target material itself. The importance to mitigate this background source has been demonstrated by several experiments, like GERDA \cite{Bar06}, CUORE \cite{CUORE08}, KamLAND-Zen \cite{kam12}, and SNO+ \cite{snop15}.

In this article we will focus on identifying the potential background nuclides for the measurement of the double-beta half-life of $^{124}$Sn in a tin-loaded liquid-scintillator based experiment. A detailed description of the method used to select the candidate nuclides and to calculate the expected production rates is given in \cite{lozza15}. The neutron and proton flux parametrization at sea-level of Armstrong and Gehrels \cite{arm73, geh85} is used in the range from 10\,MeV to 100\,GeV, while the excitation functions are based on the semi-empirical formulae of Tsao and Silberberg \cite{silb73, silb77} for E$>$ 100\,MeV as implemented in the ACTIVIA code \cite{activia}. For energies in the range 10\,MeV$<$E$<$200\,MeV the TENDL-2012 library for neutrons and protons \cite{talys} is used if available. 

\subsection{Nuclides}\label{sec::nuclides}

The main goal of the proposed tin experiment is the measurement of the 2 $\nu$ double-beta decay half-life of $^{124}$Sn, which requires the identification of a good fraction of the double-beta decay spectrum, which is sufficient as the spectral shape of the spectrum is theoretically well known. For the purpose of this paper we will assume a minimum energy of 600\,keV, and an energy resolution of 6\%/$\sqrt{E}$ at 1\,MeV. This value is aligned with what has been found by other loaded liquid-scintillator based experiments. It is expected that for energies smaller than 600\,keV other backgrounds (isotopes from the U- and Th-chains) will dominate.\\ 
The criteria used for the selection of potential background nuclides in tin-loaded liquid-scintillator based experiments are the following:
\begin{itemize}
\item $Z<38$: 
   \begin{itemize}
      \item $Q>1500\,$keV with $T_{1/2}>20$\,d, or
      \item $Q>1500$\,keV with $T_{1/2}<20$\,d fed by a long lived parent.
   \end{itemize}
These nuclides are produced by spallation reactions of high energy protons and neutrons and have generally a small production rate. Shorter-lived isotopes are usually removed by a cooling down time of about 6 months at an underground location (like the Gran Sasso National Laboratory (LNGS) in Italy or SNOLAB in Canada).
\item $38\leq Z\leq51$: 
\begin{itemize}
\item $Q>600\,$keV with $T_{1/2}>5$\,d, or
\item $Q>600$\,keV with  $T_{1/2}<5$\,d fed by a long lived parent.
\end{itemize}
These are the nuclides close in mass number to Sn and that can have a high production rate even for low energy neutrons and protons. Therefore, short-lived isotopes can still be present after a cooling-down time of 6 months. Depending on the duration of the  trigger window of the experiment (time window in which events are collected) summing effects and pile-up due to high background rates could also be a potential background.
\end{itemize}

A list of potential background candidates is given in Table \ref{tab::bckg_cand} along with the Q-value, the half-life, and the decay-mode. The expected dominant backgrounds for the measurement of the T$_{1/2}^{2\beta 2\nu}$($^{124}$Sn) are isotopes of Sn, Sb, Cd, Rh, In, and Y.

\begin{table*}[ht]\centering
\scalebox{0.89}{
\footnotesize
\begin{tabular}{l|lcl}
Isotope & $T_{1/2}$ & $Q$-value  [MeV] & Decay mode (BR [\%]) \\ \hline
$^{125}$Sb ($^{125}$Sn and direct) &2.76\,yr (9.64\,d)  &0.767&$\beta^-$ \\ 
$^{124}$Sb                      &60.2\,d                &2.90&$\beta^-$ \\ 
$^{120m}$Sb              &5.76\,d                &2.68&EC \\
$^{125}$Sn                      &9.64\,d                &2.36&$\beta^-$ \\ 
$^{123}$Sn                      &129\,d                 &1.41&$\beta^-$ \\ 
$^{113}$Sn                      &115\,d                 &1.04&EC \\ 
$^{114}$In($^{114m}$In)  &71.9\,s (49.5\,d)     &1.45&EC, $\beta^+$ (0.5)\\ 
$^{114}$In($^{114m}$In)  &71.9\,s (49.5\,d)     &2.0& $\beta^-$ (99.5)\\ 
$^{115m}$Cd              &44.6\,d                &1.63&$\beta^-$ \\ 
$^{110}$Ag ($^{110m}$Ag) &24.6\,s (250\,d)       &2.89&$\beta^-$ (99.7) \\ 
$^{110}$Ag ($^{110m}$Ag) &24.6\,s (250\,d)       &0.889&EC (0.3)\\ 
$^{110m}$Ag & 250\,d      & 3.0 &$\beta^-$ (98.6) \\ 
$^{108}$Ag ($^{108m}$Ag) &2.38\,min (438\,yr)    &1.65&$\beta^-$ (97.15)\\ 
$^{108}$Ag ($^{108m}$Ag) &2.38\,min (438\,yr)    &1.92&EC, $\beta^+$ (2.85) \\ 
$^{108m}$Ag              &438\,yr                &2.03&EC (91.3)\\
$^{106m}$Ag              &8.28\,d                &3.05&EC \\
$^{105}$Ag                      &41.3\,d                &1.35&EC\\ 
$^{106}$Rh ($^{106}$Ru)      &30.1\,s (372\,d)       &3.54&$\beta^-$ \\
$^{102}$Rh                      &207\,d                 &2.32&EC, $\beta^+$ (78) \\ 
$^{102}$Rh ($^{102m}$Rh and direct)&207\,d (3.74\,y) &1.15&$\beta^-$ (22)\\ 
$^{102m}$Rh              &3.74\,yr               &2.46&EC (99.767) \\ 
$^{103}$Ru                      &39.2\,d                &0.764&$\beta^-$ \\ 
$^{98}$Tc                       &4.20$\cdot10^{6}$\,yr  &1.79&$\beta^-$ \\ 
$^{95}$Tc ($^{95m}$Tc)   &20.0\,h (61.0\,d)        &1.69&EC \\
$^{95m}$Tc               &61.0\,d                  &1.73&EC, $\beta^+$ (96.12) \\ 
$^{95}$Nb ($^{95}$Zr)           &35.0\,d (64.0\,d)      &0.926&$\beta^-$\\ 
$^{94}$Nb                       &20300\,yr              &2.04&$\beta^-$ \\
$^{92}$Nb                       &3.47$\cdot10^{7}$\,yr  &2.01&EC \\
$^{91}$Nb ($^{91m}$Nb and direct)    &680\,yr (60.7\,d) &1.26&EC, $\beta^+$  \\ 
$^{91m}$Nb               &60.9\,d                &1.36&EC, $\beta^+$ (3.4)\\ 
$^{95}$Zr                       &64.0\,d                &1.12&$\beta^-$ \\ 
$^{88}$Zr                       &83.4\,d                &0.67&EC \\ 
$^{91}$Y                        &58.5\,d                &1.54&$\beta^-$ \\ 
$^{90}$Y ($^{90}$Sr)            &3.19\,h (28.9\,yr)      &2.28&$\beta^-$ \\  
$^{89}$Sr                       &50.5\,d                &1.50&$\beta^-$\\ 
$^{85}$Sr                       &64.9\,d                &1.06&EC \\ 
$^{82}$Rb ($^{82}$Sr)           &1.26\,min (25.3\,d)    &4.40&EC, $\beta^+$  \\ 
$^{84}$Rb                       &32.8\,d                &2.68&EC, $\beta^+$ (96.1) \\
$^{68}$Ga ($^{68}$Ge)           &1.13\,h (271\,d)       &2.92&EC, $\beta^+$ \\ 
$^{60}$Co ($^{60}$Fe and direct) &1930\,d (2.62$\cdot10^6\,$yr)&2.82&$\beta^-$ \\
$^{58}$Co                       &70.9\,d                &2.31&EC, $\beta^+$\\ 
$^{56}$Co                       &77.2\,d                &4.57&EC, $\beta^+$ \\
$^{59}$Fe                       &44.5\,d                &1.56&$\beta^-$ \\ 
$^{46}$Sc                       &83.8\,d                &2.37&$\beta^-$ \\ 
$^{44}$Sc ($^{44}$Ti)           &3.97\,h (60.0\,yr)     &3.65&EC, $\beta^+$ \\ 
$^{42}$K ($^{42}$Ar)            &12.3\,h (32.9\,yr)     &3.53&$\beta^-$ \\ 
$^{40}$K                        &1.25$\cdot10^{9}$\,yr  &1.50&EC, $\beta^+$ (10.72)\\ 
$^{32}$P ($^{32}$Si)            &14.3\,d (153\,yr)      &1.71&$\beta^-$\\
$^{26}$Al                       &7.17$\cdot10^{5}$\,yr  &4.00&EC, $\beta^+$  \\
$^{22}$Na                       &2.60\,yr               &2.84&EC, $\beta^+$ \\ \hline
\end{tabular}}
\caption{Potential cosmogenic-induced background candidates for the measurement of T$_{1/2}^{2\beta 2\nu}$ of $^{124}$Sn. Shown is the T$_{1/2}$, the Q-value, and the decay-mode with the respective branching ratio (BR). Values are from \cite{Nudat}. When a short-lived nuclide is fed by a long-lived one its T$_{1/2}$ is given in brackets. \label{tab::bckg_cand}}
\end{table*}

\subsection{Expected background rates}

The production rates for the various nuclides have been calculated using the input parameters described at the beginning of this section and in \cite{lozza15}, and are shown in column two of Table \ref{tab::prod}. A natural tin target has been assumed. The contribution from short-lived parents (fraction of seconds to seconds of half-life) to the various nuclides' production rates has been included. For the isotopes $^{113}$Sn, $^{117m}$Sn, $^{119m}$Sn, $^{121m}$Sn, $^{123}$Sn, and $^{125}$Sn the production mode via thermal neutron capture is also included. The thermal neutron cross sections are given in \cite{thermxs, kra06, mir97, nel50}, while the flux on surface is from \cite{thermal}. Among the various isotopes produced the most dangerous one is $^{123}$Sn which has both a long half-life of 129 days and a high Q-value of 1.41\,MeV.

Starting from the production rates given in column two of Table \ref{tab::prod}, the expected activity and number of events in a year of data taking are calculated for two different scenarios and are also shown in Table \ref{tab::prod} columns three to six. The activity and the numbers of events are calculated using the formula (1) in \cite{lozza15}. The assumed scenarios are:
\begin{description}
\item[Scenario 1] An exposure at sea-level of 1 year, followed by a 6 months of cooling down time deep underground (UG).
\item[Scenario 2] An exposure at sea-level of 1 year, followed by a purification factor (for instance via distillation) of 10$^{3}$, common to all the nuclides with the exception of the Sn isotopes. It is assumed that the purification will happen on surface, and therefore a re-activation period of 10 days before bringing the material underground is added. A final 6 or 12 months cooling-down period deep UG, before data taking, is added.
\end{description}

It is assumed that the activation due to the deep underground neutron flux induced by muons and ($\alpha$,n) reactions from the rock is negligible.

From Table \ref{tab::prod} it is clear that the major contribution to the background arises from $^{113}$Sn, $^{123}$Sn, $^{125}$Sb, $^{102}$Rh, $^{114}$In, $^{102m}$Rh, $^{88}$Zr, and $^{105}$Ag.\\

\textbf{$^{113}$Sn: }\\
$^{113}$Sn decays by electron capture (EC) emitting two main gammas of 0.65\,MeV (BR = 2.21\%) and 0.39\,MeV (BR = 97.79\%). For a liquid-scintillator based experiment with an energy resolution of 6\%$/\sqrt{E}$ it will
be challenging to measure this isotope in the low energy part of the spectrum. Since this isotope might not be reduced by purification, the mitigation strategy consists of  short exposure times on surface and long cooling-down times UG.\\

\textbf{$^{123}$Sn:}\\
$^{123}$Sn $\beta^{-}$decays ($Q=1.408$\,MeV) to the ground state of $^{123}$Sb (BR = 99.37\%). This is the major background source for the measurement of the half-life of $^{124}$Sn, since it might not be reduced by purification. One year of activation on surface followed by a year of cooling-down time results in about 200 events in the first year of data taking per kg of Sn. For comparison, for a theoretical T$_{1/2}^{2\beta 2\nu}$ of 1.5$\cdot 10^{21}$ yr for $^{124}$Sn, the expected number of decays is 130 events/yr/kg of Sn. It is therefore important to minimize as much as possible the exposure on surface and to maximize the time spent UG. The region above about 1.5 MeV is not effected by this isotope.\\

\textbf{$^{125}$Sb: }\\
$^{125}$Sb $\beta^{-}$-decays with a $Q$-value of 0.77\,MeV. The resulting $\beta+\gamma$ spectrum can hide the low energy part of the $\beta\beta$ spectrum of $^{124}$Sn. The isotope is produced both directly and via the decay of $^{125}$Sn. Thanks to the relative short half-life of $^{125}$Sn (see Tab. \ref{tab::bckg_cand}), the activity of this parent isotope is highly reduced with a 6 months cooling-down period.\\

\textbf{$^{102}$Rh:}\\
$^{102}$Rh decays by EC (BR = 63.3\%), $\beta^{-}$ (BR = 22\%), and $\beta^{+}$ (BR = 11.7\%). The $\beta^{-}$ decay has a $Q$-value of 1.150\,MeV, while the $\beta^{+}$ channel decays in 10.5\% of the cases directly to the ground state of  $^{102}$Ru with a $Q$-value of $2.32$\,MeV. One year of exposure at the sea-level followed by six months cooling-down time deep UG will result is 100 events/yr/kg of Sn, compared to the 130 events/yr/kg expected for $^{124}$Sn. A mitigation strategy which includes a purification factor of at least 100 and long cooling down time UG will effectively reduce this background source.\\

\textbf{$^{102m}$Rh:}\\
$^{102m}$Rh decays by EC (BR = 99,767\%) to the ground state of  $^{102}$Ru with a Q-value of 2.5\,MeV. The main gamma line of  2.22\,MeV (67\%), and 1.87\,MeV (35\%), can be challenging for the measurement of the $^{124}$Sn half-life. One year of exposure at the sea-level followed by six months cooling-down time deep UG will result is 100 events/yr/kg of Sn, compared to the 130 events/yr/kg expected for $^{124}$Sn. Due to the long half-life of 3.74 yr, the mitigation strategy should include a purification factor of at least 100 followed by a short re-activation time, ideally of a maximum of 1 day.\\

\textbf{$^{114}$In:}\\
With a $Q$-value of $1.99$\,MeV the $\beta^{-}$-decay (BR = 99.5\%) of $^{114}$In is also covering a good part of the $\beta\beta$ spectrum of $^{124}$Sn. However, due to the relative short half-life of about 50 days this isotope decays away relatively fast. \\

In summary under the second scenario with 6 months of cooling down, a total of 1950 cosmogenic-induced background events/yr/kg of Sn are expected. Of these events more than 97\% are due to $^{113}$Sn (69\%), $^{123}$Sn (28\%), and $^{102m}$Rh (0.1\%). In order to effectively reduce these backgrounds a maximum re-activation time of 1 day after purification, together with cooling-down time larger than 2 years would be necessary.

\subsection{Enriched tin target}

The use of a 99.9\% $^{124}$Sn enriched target would reduce the contribution of non-Sn background nuclides by about a factor 2, while the number of expected signal events will increase by about a factor 20. This would allow a longer re-activation period at sea-level following the purification process. On the other hand the background due to Sn isotopes, in particular $^{123}$Sn, will increase more or less by the same amount.

%Another important parameter that needs to be considered when selecting the site for the tin-loaded experiment is the thermal neutron activation that can happen while the material is stored UG, considering that the major backgrounds are Sn isotopes.

\begin{table*}[ht]\centering
\scalebox{0.85}{
\footnotesize
\begin{tabular}{cccccc}
  \hline
Isotope 	&	 \textit{R}  ($\phi$ from \cite{arm73}\cite{geh85}) & \multicolumn{2}{c}{Scenario 1} & \multicolumn{2}{c}{Scenario 2} \\ \cline{3-4} \cline{5-6}
	&	[$\mu$Bq/kg]	&	A [Bq/kg] & Events/yr/kg & A [Bq/kg] & Events/yr/kg \\\hline
$^{125}$Sb ($^{125}$Sn and direct) 	&	0.03	&	8.15E-6	&	227.31	&	6.07E-7	&	16.94	\\
$^{124}$Sb                      	&	0.58	&	6.98E-8	&	0.52	&	7.77E-9	&	0.06	\\
$^{120m}$Sb              	&	5.46	&	1.56E-15	&	1.12E-9	&	1.09E-15	&	7.84E-10	\\
$^{125}$Sn                      	&	41.15	&	8.16E-11	&	9.80E-5	&	8.16E-11	&	9.80E-5	\\
$^{123}$Sn                      	&	122.44	&	3.95E-5	&	546.29	&	3.98E-5	&	550.97	\\
$^{113}$Sn                      	&	351.53	&	1.04E-4	&	1327.4	&	1.05E-4	&	1337.0	\\
$^{114}$In($^{114m}$In) beta-  	&	(256.33)	&	1.91E-5	&	116.68	&	2.53E-6	&	15.44	\\
$^{114}$In($^{114m}$In) EC  	&	(256.33)	&	1.91E-5	&	0.59	&	2.53E-6	&	0.08	\\
$^{115m}$Cd              	&	27.86	&	1.62E-6	&	8.97	&	2.36E-7	&	1.30	\\
$^{110}$Ag ($^{110m}$Ag) beta-	&	(6.51)	&	3.40E-8	&	0.67	&	1.49E-9	&	0.03	\\
$^{110}$Ag ($^{110m}$Ag) EC	&	(6.51)	&	3.40E-8	&	2.02E-3	&	1.49E-9	&	8.88E-5	\\
$^{110m}$Ag 	&	6.51	&	2.50E-6   &  48.88	&	1.10E-7  &   2.15	\\
$^{108}$Ag ($^{108m}$Ag) beta- 	&	(22.57)	&	3.10E-9	&	0.10	&	8.81E-11	&	2.70E-3	\\
$^{108}$Ag ($^{108m}$Ag) EC/beta+	&	(22.57)	&	3.10E-9	&	2.84E-3	&	8.81E-11	&	8.06E-5	\\
$^{108m}$Ag              	&	22.57	&	3.57E-8	&	1.03	&	1.01E-9	&	0.03	\\
$^{106m}$Ag              	&	60.43	&	1.39E-11	&	1.31E-5	&	7.86E-12	&	7.41E-6	\\
$^{105}$Ag                      	&	118.65	&	5.52E-6	&	28.34	&	8.59E-7	&	4.41	\\
$^{106}$Rh ($^{106}$Ru)      	&	(0.08)	&	2.85E-8	&	0.65	&	1.10E-9	&	0.03	\\
$^{102}$Rh ($^{102m}$Rh and direct) EC                 	&	18.49 (19.28)	&	7.08E-6	&	100.60	&	3.37E-7	&	4.79	\\
$^{102}$Rh ($^{102m}$Rh and direct) beta-	&	18.49	 (19.28) &	7.08E-6	&	28.41	&	3.37E-7	&	1.35	\\
$^{102m}$Rh              	&	19.28	&	2.97E-6	&	85.44	&	9.19E-8	&	2.64	\\
$^{103}$Ru                      	&	1.24	&	4.90E-8	&	2.40E-1	&	7.99E-9	&	0.04	\\
$^{98}$Tc                       	&	5.47	&	9.03E-13	&	2.85E-5	&	2.56E-14	&	8.09E-7	\\
$^{95}$Tc ($^{95m}$Tc)   	&	(21.53)	&	1.05E-7	&	0.78	&	1.15E-8	&	0.09	\\
$^{95m}$Tc               	&	21.53	&	2.66E-6	&	19.13	&	2.93E-7	&	2.10	\\
$^{95}$Nb ($^{95}$Zr)           	&	0.96	&	6.79E-8	&	0.46	&	8.71E-9	&	0.06	\\
$^{94}$Nb                       	&	1.90	&	6.48E-11	&	2.05E-3	&	1.84E-12	&	5.81E-5	\\
$^{92}$Nb                       	&	8.27	&	1.65E-13	&	5.21E-6	&	4.69E-15	&	1.48E-7	\\
$^{91}$Nb ($^{91m}$Nb and direct)    	&	29.98	&	4.54E-8	&	1.44	&	1.25E-9	&	0.04	\\
$^{91m}$Nb               	&	15.53	&	1.91E-6	&	0.48	&	2.11E-7	&	0.05	\\
$^{95}$Zr                       	&	0.16	&	2.17E-8	&	0.17	&	2.29E-9	&	0.02	\\
$^{88}$Zr                       	&	21.08	&	4.40E-6	&	43.53	&	3.73E-7	&	3.69	\\
$^{91}$Y                        	&	0.33	&	3.70E-8	&	0.27	&	4.22E-9	&	0.03	\\
$^{90}$Y ($^{90}$Sr)            	&	(0.08)	&	1.89E-9	&	0.06	&	5.42E-11	&	1.69E-3	\\
$^{89}$Sr                       	&	0.17	&	1.34E-8	&	0.08	&	1.74E-9	&	1.09E-2	\\
$^{85}$Sr                       	&	17.82	&	2.48E-6	&	19.64	&	2.59E-7	&	2.05	\\
$^{82}$Rb ($^{82}$Sr)           	&	(8.11)	&	5.49E-8	&	0.17	&	1.32E-8	&	0.04	\\
$^{84}$Rb                       	&	1.44	&	3.05E-8	&	0.12	&	5.83E-9	&	0.02	\\
$^{68}$Ga ($^{68}$Ge)           	&	(2.28)	&	8.66E-7	&	17.76	&	3.69E-8	&	0.76	\\
$^{60}$Co ($^{60}$Fe and direct) 	&	1.31 (0.24)	&	1.51E-7	&	4.47	&	4.56E-9	&	0.13	\\
$^{58}$Co                       	&	3.34	&	5.44E-7	&	4.67	&	5.27E-8	&	0.45	\\
$^{56}$Co                       	&	0.43	&	7.97E-8	&	0.74	&	7.18E-9	&	0.07	\\
$^{59}$Fe                       	&	0.47	&	2.72E-8	&	0.15	&	3.96E-9	&	0.02	\\
$^{46}$Sc                       	&	1.75	&	3.68E-7	&	3.65	&	3.10E-8	&	0.31	\\
$^{44}$Sc ($^{44}$Ti)           	&	(0.093)	&	1.06E-9	&	0.03	&	3.03E-11	&	9.51E-4	\\
$^{42}$K ($^{42}$Ar)            	&	(0.16)	&	1.96E-8	&	0.54	&	6.32E-10	&	0.02	\\
$^{40}$K                        	&	2.21	&	1.23E-15	&	4.15E-9	&	3.49E-17	&	1.18E-10	\\
$^{32}$P ($^{32}$Si)            	&	2.71 (0.66)	&	3.36E-9	&	9.46E-2	&	2.31E-10	&	2.93E-3	\\
$^{26}$Al                       	&	0.87	&	8.41E-13	&	2.65E-5	&	2.39E-14	&	7.53E-7	\\
$^{22}$Na                       	&	1.25	&	2.56E-7	&	7.10	&	8.21E-9	&	0.23	\\
 \hline
\end{tabular}}
\caption{Expected production rate (\textit{R}) for the isotopes listed in Table \ref{tab::bckg_cand}. The values are obtained using the ACTIVIA code \cite{activia} for \textit{E}$>$100\,MeV with an energy step of 10\,MeV. If available the cross sections for 10\,MeV$<$\textit{E}$<$200\,MeV are obtained from the TENDL-2012 library \cite{talys} with an energy step of 1 keV. The flux parametrization is the one from Armstrong and Gehrels \cite{arm73}\cite{geh85}. The decay branching ratios are taken into account. The two different scenarios are outlined in the text. The production rates of the parent nuclides are shown in brackets.}
\label{tab::prod}
\end{table*}

\section{Sensitivity}
The classic choice for such an experiment would be a copy of experiments like Borexino, SNO+, 
Daya Bay and other experiments in a way to fill the Sn-compound in a liquid scintillator sphere readout by photomultipliers.
A fiducial volume of about 100 liters of tin-loaded scintillator surrounded by pure scintillator, separated by a 
balloon or acrylic sphere could be an option. The latter would reduce external background
from the walls and phototubes significantly. If the loading of 20 kg of tin into LAB (which is minor compared to the SNO+ experiment planning
to load more than a ton of Te into LAB) is performed, a rate of 2600 decays/yr would result for a half-life of 1.5$\times10^{21}$ yrs . 
Depending on the background and accumulation of sufficient statistics for a reasonably precise half-life determination, the experiment
might has to run for some years. 

\section{Summary and conclusions}
This paper discussed a potential measurement of the \zbb decay of \tinhv, assuming
that half-lives could be as been as long as 10$^{22}$ years. The approach would be a tin-loaded
scintillator readout by photomultipliers and stored in a deep underground experimental laboratory.
Optical properties and the long term stability of the considered compound have been measured 
and in addition a detailed estimation of cosmogenic backgrounds have been performed. The most
critical isotopes have been identified and require, together with a mitigation strategy based on a long cooling down period UG, a certain time 
before the experiment can be performed. This can only be shortened if the opportunity exists
to get freshly produced tin with minimal surface time between factory and laboratory or if the purification can be performed UG.

\section{Acknowledgement}
We acknowledge the support of the Deutsche Akademischer Austauschdienst (DAAD-Contract number 56012129)

\end{document}